\documentclass[twocolumn,aps,prl,10pt]{revtex4}
\usepackage[dvips]{graphicx}

\begin{document}
\title{Comment on Phys. Rev. Lett.'s paper "All-Electron Self-Consistent GW Approximation: Application to Si, MnO, and NiO":
\\
band {\it vs} localized description of NiO}
\author{R. J. Radwanski}
\affiliation{Center of Solid State Physics, S$^{nt}$Filip 5,
31-150 Krakow, Poland,
\\
Institute of Physics, Pedagogical University, 30-084 Krakow,
Poland
\\
http:www.css-physics.edu.pl; E-mail: sfradwan@cyf-kr.edu.pl}

\author{Z. Ropka} \affiliation{Center of Solid State Physics,
S$^{nt}$Filip 5, 31-150 Krakow, Poland}

\begin{abstract}
In contrary to authors of Phys. Rev. Lett. {\bf 93}, 126406
(2004) claiming "the band picture to be a reasonable starting
point for the description of the electronic structure of NiO,
much better than the ligand-field picture", we argue that the
many-electron CEF approach is physically adequate starting point
for discussion of the electronic structure and magnetism of NiO.

\end{abstract}

\pacs{75.50.Ee, 75.10.Lp, 71.70.Ej} \keywords{orbital magnetic
moment, NiO, spin-orbit coupling, crystal field}

\maketitle

\vspace {-0.01 cm} By this Comment we would like to express our
large disagreement about a conclusive statement of a paper
\cite{1} in Phys. Rev. Lett. {\bf 93}, 126406 (2004) that authors
Faleev, van Schilfgaarde, and Kotani "believe that the band
picture [23, here \cite{2}] for NiO is a reasonable starting
point for the description of the electronic structure of NiO,
much better than previously thought, and in many respects more
appropriate than the ligand-field picture." Faleev et al.
\cite{1} have performed electronic-structure calculations with "a
new kind of self-consistent GW (SCGW) approximation based on the
all-electron, full potential linear muffin-tin orbital method."
getting continuous energy spectrum, a band of $d$ electrons wide
by 5-6 eV with a spin-polarization of $e_{g}$-$e_{g}$ states by
2-3 eV similar to that got by Terakura et al. \cite{2,3}. In
contrary to authors of Ref. \cite{1} we argue that the
many-electron crystal-field approach, known from works of Bethe
and Van Vleck from 1929 with predominantly Ni$^{2+}$ and O$^{2-}$
ions, is physically adequate starting point for discussion of the
electronic structure and magnetism of NiO \cite{4,5,6}. Being
more exact, we claim that the continuous electronic structure for
$d$ electrons presented in Figs 2 and 3 of Ref. \cite{1} is not
realized in the reality - we claim that in NiO exists the discrete
electronic structure.

\vspace {-0.01 cm} In our previous Comment \cite{7} to this paper
we put attention that the "excellent agreement in the SCGW
approach" of the calculated magnetic moment of 1.72 $\mu _{B}$ is
based on comparison to too low value derived in 1983 but not to a
value of 2.2$\pm $0.3 $\mu _{B}$ at 300 K  derived in 1998
\cite{8,9}. The disagreement becomes larger if one compares a more
relevant zero-temperature value of 2.6 $\mu _{B}$.

The description of the electronic structure and magnetism of 3$d$
oxides, including famous NiO, is a subject of extensive
theoretical and experimental investigations by last 70 years. In
recent years the theoretical understanding becomes a subject of
enormous controversy when the localized-based view is openly
discriminated in Phys. Rev. Lett. and Phys. Rev. B \cite{10}. It
is a direct reason for writing of this Comment by us as the
publication of Comment is the scientific obligation of each
journal considered itself to be scientific. It is obvious that
Physics can develop only in the open integral scientific
discussion and in the friendly atmosphere, but it is also obvious
that in this discrimination takes part great modern physicists
acting as referees. This discrimination is the best proof that
the most natural and the most elegant approach as the CEF theory
is is not appreciated in the theory of the solid-state physics at
the beginning of the XXI century. Despite it we consequently
develop an understanding of transition-metal solids with the
discrete electronic structure leaving its solution to the future
generation of physicists.

Although the ionic picture is known almost from the beginning of
the modern magnetism, in particular from the seminal works of
Bethe and Van Vleck on the crystal-field theory and of Tanabe and
Sugano on the effect of the octahedral crystal field on 3$d$
electronic terms we have extended the single-site CEF theory to
the Quantum Atomistic Solid State Theory (QUASST) for 3$d$
compounds \cite{11,12,13} by the correct treatment of the
spin-orbit coupling and the recognition that the many-electron
CEF approach itself incorporates strong electron correlations
among $d$ electrons already from the beginning. We start analysis
of NiO from the detailed analysis of the single-ion effects and
the evaluation of the discrete low-energy electronic structure of
the Ni$^{2+}$ ion. After the completion with inter-site
spin-dependent interactions we have calculated the
magnetically-ordered state with the Ni magnetic moment at T = 0 K
as 2.54 $\mu _{B}$. QUASST allows calculations of the orbital and
spin moment as well as physically adequate thermodynamics
\cite{4}. The insulating state occurs both in the
antiferromagnetic and paramagnetic state. We describe in the
consistent way both the paramagnetic and the magnetic state with
the description of, for instance, the $\lambda$-peak at $T_{N}$
in the temperature dependence of the heat capacity \cite{13,14}.

QUASST can be practically understood as a demand to start from the
very strong electron correlation limit and to evaluate
crystal-field interactions the first. It is a surprise that
despite long-time studies the basic interactions, like exchange
splitting, the ligand-field, the covalency, the hybridization and
intra- and inter-sublattice coupling, discussed in the band
picture of Refs \cite{1,2,3}, have not been quantified yet. In
Ref. \cite{1} they are considered in the eV accuracy. In QUASST
we have show that the details of the electronic structure in
scale of 1 meV are important. Namely, the trigonal off-octahedral
distortion of the 1 meV effect determines the direction of the Ni
magnetic moment.

\vspace {-0.01 cm}In conclusion, we claim that the ionic-based
approach with localized $d$ electrons and with discrete energy
states in scale of 1 meV as are discussed in many-electron
crystal-field theory is the physically adequate starting point
for consideration of the magnetism and electronic structure of
NiO. We claim that any approach neglecting very strong
correlations among $d$ electrons, the spin-orbit coupling and the
orbital magnetism is not physically adequate to 3$d$ oxides, in
particular not to NiO.

\end{document}